# Modeling the "rapid" part of the velocity – pressure-gradient correlation in inhomogeneous turbulent flows

Svetlana V. Poroseva

School of Computational Science, Florida State University, FL 32306

A new model for the "rapid" part of the velocity – pressure-gradient correlation in the Reynolds averaged Navier-Stokes equations is suggested. It is shown that in an inhomogeneous incompressible turbulent flow, the model that is linear in the Reynolds stresses should include two model coefficients. A functional form of the coefficients is analyzed in some extreme turbulence states. As demonstrated, the proposed model reduces to the standard linear pressure-strain model in homogeneous turbulence and can satisfy realizability conditions provided the coefficients are functions of the mean velocity gradients and the Reynolds stresses. In an inhomogeneous turbulent flow, coefficients should depend on parameters directly related to inhomogeneous effects. The model is validated using direct numerical simulation data from two flows: the plane wake and the plane mixing layer.

# 1. Introduction

The exact transport equation for the Reynolds stresses provides more detailed information on turbulence structure than the transport equation for the turbulent kinetic energy combined with the Boussinesq turbulent-viscosity hypothesis for the Reynolds stresses. However, the exact Reynolds stress transport (RST) equation

$$\frac{\partial <u_i u_j>}{\partial t} + U_k <u_i u_j>_{,k} = P_{ij} - <u_i u_j u_k>_{,k} + \Pi_{ij} + \nu \nabla^2 <u_i u_j> - \varepsilon_{ij}, \quad (1.1)$$

cannot be solved directly. The terms: $<u_i u_j u_k>_{,k}$ (turbulent diffusion), $\varepsilon_{ij} = 2\nu \langle u_{i,k} u_{j,k} \rangle$ (dissipation tensor), and $\Pi_{ij} = -\frac{1}{\rho}(<u_i p_{,j}> + <u_j p_{,i}>)$ (velocity – pressure-gradient tensor), are unknown and require modeling. The following notation is used in (1.1): $P_{ij} = - <u_i u_k> U_{j,k} - <u_j u_k> U_{i,k}$; $U_i$ and $u_i$ are the mean and fluctuating velocity components; $<...>$ means ensemble average; $-\rho <u_i u_j>$ are the Reynolds stresses, $\rho$ is the density, $p$ is the pressure fluctuation, and $\nu$ is the kinematic viscosity. For the sake of simplicity, Cartesian tensor notation is used in (1.1) and in what follows. In this notation $f_{i,j} = \partial f_i / \partial x_j$. Clearly, the ability of a turbulence model based on the exact RST equation to describe flow physics depends on the quality of model expressions incorporated in the RST equation to represent unknown terms. The focus of the current study is on modeling the velocity – pressure-gradient tensor $\Pi_{ij}$.

Following Rotta (1951), it is a common practice to split the correlation $<u_i p_{,j}>$ into two parts: the pressure-strain correlation $<u_{i,j} p>$ and the pressure-diffusion part $<u_i p>_{,j}$. This approach has advantages if one simulates homogeneous turbulence with a two-equation turbulence model. In homogeneous turbulence, both the pressure-diffusion term $<u_i p>_{,i}$ and the pressure-strain term $<u_{i,i} p>$ do not contribute to the transport of the turbulent kinetic energy. In inhomogeneous turbulent flows, however, the modeling of the pressure-diffusion term in the transport equation for the turbulent kinetic energy is challenging. Direct numerical simulation (DNS) data from free shear flows (Rogers & Moser 1994, Moser, Rogers & Ewing 1998) show that the contribution of the pressure diffusion to the turbulent kinetic energy balance is not negligible, especially in the central core of these flows. These data also show that modeling the pressure diffusion and turbulent diffusion terms together is not likely to be successful because they have qualitatively different profiles (see Fig. 21 in Rogers & Moser 1994 and Fig. 12 in Moser, Rogers & Ewing 1998). In fact, a model for turbulent diffusion can absorb only a part of the pressure-diffusion term $<u_i p>_{,i}$, the so-called "slow" part:

$$-\frac{1}{\rho}<u_i p>_{,i}^{(s)} = \frac{1}{5}<u_m u_m u_i>_{,i} \qquad (1.2)$$

(Lumley 1978). The rest of the pressure-diffusion term still requires modeling.

The above discussion demonstrates that the splitting of the velocity – pressure-gradient correlation $<u_i p_{,j}>$, which is originally present in equation (1.1), into the pressure-strain and pressure-diffusion parts is not so beneficial in inhomogeneous turbulent flows as in homogeneous turbulence, even in application to the transport equation for the turbulent kinetic energy. In regard to equation (1.1), both terms $<u_i p>_{,j}$ and $<u_{i,j} p>$ should be modeled based on DNS data. Thus, two models for the pressure-related correlations are necessary if the splitting of $<u_i p_{,j}>$ is imposed. This is clearly a disadvantage. Therefore, in inhomogeneous turbulent flows, modeling the correlation $<u_i p_{,j}>$ is more natural.

The exact integral expression for this correlation

$$-\frac{1}{\rho}<u_i p_{,j}> = -\frac{1}{2\pi}\iiint \left[U'_{m,n}<u'_n u_i>_{,m}\right]'_{,j}\frac{1}{r}dV' - \frac{1}{4\pi}\iiint <u'_m u'_n u_i>'_{mnj}\frac{1}{r}dV'$$
$$-\frac{1}{4\pi}\iint \left\{\frac{1}{r}\frac{\partial <p'_{,j}u_i>}{\partial n'} - <p'_{,j}u_i>\frac{\partial}{\partial n'}\left(\frac{1}{r}\right)\right\}dS' \qquad (1.3)$$

(Chou 1945) contains three terms. In expression (1.3), " ' " above a flow variable indicates that it should be evaluated at a point $Y'$ with coordinates $x'_i$, which ranges over the region of the flow; $r$ is the distance from $Y'$ to the point $Y$ with coordinates $x_i$; $dV'$ and $dS'$ are the volume and surface elements, respectively; and $\partial/\partial n'$ denotes the normal derivative. The velocity – pressure-gradient correlation on the left side of (1.3) is evaluated at point $Y$, whereas all derivatives on the right side are taken at $Y'$. The objective of the current paper is to develop a model for the first term on the right side of expression (1.3), which is called the "rapid" part of the velocity – pressure-gradient correlation $<u_i p_{,j}>$ due to its relation to the mean velocity gradient.

## 2. Modeling the "rapid" part of the velocity – pressure-gradient correlation

The integral expression for the "rapid" part of the velocity – pressure-gradient correlation

$$-\frac{1}{\rho}<u_i p_{,j}>^{(r)} = -\frac{1}{2\pi}\iiint \left[U'_{m,n}<u'_n u_i>'_{,m}\right]_{,j}\frac{1}{r}dV' \quad (2.1)$$

holds exactly in an incompressible flow. In a compressible flow, it holds approximately. The integrand in (2.1) is non-zero only over the volume where the two-point correlation $<u'_n u_i>$ (or more precisely $<u'_{n,m} u_i>$) does not vanish. In other words, for a fixed point $Y$, only those points $Y'$ which lie within the length scale of the two-point correlation measured from $Y$ contribute to the integral in (2.1). If one assumes that the function $U'_{m,n}$ varies more slowly than the two-point correlation within the volume determined by the length scale of the two-point correlation, then, to a first approximation, we can rewrite expression (2.1) as

$$-\frac{1}{\rho}<u_i p_{,j}>^{(r)} = -\frac{1}{2\pi}U_{m,n}\iiint \left[<u'_n u_i>'_{,m}\right]_{,j}\frac{1}{r}dV' \quad (2.2)$$

(Chou 1945). This is the so-called "weak inhomogeneity" approximation wherein the mean velocity gradient is assumed "almost constant" over an integral length scale and all functions are still regarded as functions of the space coordinates. (Of course, (2.1) and (2.2) are equivalent in the case of homogeneity.) Using (2.2), Chou (1945) suggested modeling the "rapid" part of the velocity – pressure-gradient tensor as

$$\Pi_{ij}^{(r)} = -\frac{1}{\rho}(<u_i p_{,j}>^{(r)} + <u_j p_{,i}>^{(r)}) = a_{nmji}U_{m,n}. \quad (2.3)$$

However, no specific form for the tensor function $a_{nmji}$ was suggested in that work.

In the present paper, each of the correlations $<u_i p_{,j}>^{(r)}$ and $<u_j p_{,i}>^{(r)}$ will be modeled separately. That is, instead of (2.3), the model for $\Pi_{ij}^{(r)}$ is as follows

$$\Pi_{ij}^{(r)} = \left(a_{nmji} + a_{nmij}\right)U_{m,n}, \quad (2.4)$$

where

$$-\frac{1}{\rho}<u_i p_{,j}>^{(r)} = a_{nmji} U_{m,n}$$

and

$$a_{nmji} = -\frac{1}{2\pi}\iiint \left[<u'_n u_i>'_{,m}\right]_{,j} \frac{1}{r} dV'. \tag{2.5}$$

Thus, no condition on symmetry under permutation of the indices *i* and *j* is imposed on the tensor function $a_{nmji}$. The idea of modeling the pressure-related correlations separately was applied for the first time to the pressure-strain correlations $<u_{i,j} p>$ by Rotta (1951). In relation to the $<u_i p_{,j}>$-correlation, the idea was initially discussed in Poroseva (2000).

It is important to emphasize the main difference in modeling the velocity – pressure-gradient correlation $<u_i p_{,j}>^{(r)}$ and the pressure-strain correlation $<u_{i,j} p>^{(r)}$. Although in both cases, we arrive to expression (2.4) as a model representation of $\Pi_{ij}^{(r)}$ in the tensor form, the tensor functions $a_{nmji}$ are different in two cases. For $<u_i p_{,j}>^{(r)}$, the tensor function $a_{nmji}$ is given by (2.5) and for the pressure-strain correlation $<u_{i,j} p>^{(r)}$, it is given by

$$a_{nmji} = -\frac{1}{2\pi}\iiint <u'_n u_{i,j}>'_{,m} \frac{1}{r} dV'. \tag{2.6}$$

As one can notice, the spatial derivatives in two expressions are taken at different locations. It results in different properties of expressions (2.5) and (2.6).

The general model for $a_{nmji}$ given by expression (2.5) that is linear in the Reynolds stresses is as follows

$$\begin{aligned}a_{nmji} = &k \cdot (C_1 \delta_{ij}\delta_{mn} + C_2 \delta_{in}\delta_{jm} + C_3 \delta_{im}\delta_{jn}) + C_4 <u_i u_j> \delta_{mn} + C_5 <u_i u_m> \delta_{jn} \\ &+ C_6 <u_i u_n> \delta_{jm} + C_7 <u_j u_m> \delta_{in} + C_8 <u_j u_n> \delta_{im} + C_9 <u_m u_n> \delta_{ij},\end{aligned} \tag{2.7}$$

where $\delta_{ij}$ is the Kronecker symbol and *k* is the turbulent kinetic energy: $k = 1/2 <u_i u_i>$. Based on the analysis of expression (2.5), three properties of the tensor function $a_{nmji}$ can be deduced:

(i) symmetry under permutation of indices *m* and *j*;
(ii) if $m = n$, then $a_{mmji} = 0$;

(iii) if $m = j$, then $a_{njji} = 2 <u_n u_i>$.

The first property is obvious. The second property follows rigorously from the continuity equation for the two-point velocity correlation: $<u'_n u_i>'_{,n} = 0$ (Chou 1945). The third property results from Green's theorem

$$-\frac{1}{2\pi}\iiint <u'_n u_i>'_{,mm} \frac{1}{r} dV' = 2<u_n u_i> \qquad (2.8)$$

assuming that the length scale of the two-point correlation is less than the distance from the flow points under consideration to any flow boundary, and therefore, the surface integral can be neglected in (2.8). Notice again that this is not an assumption of homogeneity. The flow can be strongly inhomogeneous far from the boundaries.

Imposing three conditions i)-iii) on expression (2.7), one obtains the final model expression for the tensor function $a_{nmji}$

$$\begin{aligned} a_{nmji} = & -\frac{1}{5}\left(<u_i u_j>\delta_{mn} + <u_i u_m>\delta_{jn}\right) + \frac{4}{5}<u_i u_n>\delta_{jm} + \\ & C_1\left[\frac{1}{2}\left(<u_i u_j>\delta_{mn} + <u_i u_m>\delta_{jn}\right) + k\left(\delta_{ij}\delta_{mn} + \delta_{im}\delta_{jn}\right) + <u_i u_n>\delta_{jm}\right. \\ & \left. -<u_j u_m>\delta_{in} - 2\left(<u_i u_n>\delta_{im} + <u_m u_n>\delta_{ij}\right)\right] + \\ & C_2\left[\frac{1}{2}\left(<u_i u_j>\delta_{mn} + <u_i u_m>\delta_{jn} - <u_j u_n>\delta_{im} - <u_m u_n>\delta_{ij}\right) + \right. \\ & \left. +k\delta_{in}\delta_{jm} - \frac{3}{2}<u_j u_m>\delta_{in}\right], \end{aligned} \qquad (2.9)$$

which contains two coefficients. Substitution of (2.9) in (2.4) gives the following model for the "rapid" part of the velocity – pressure-gradient tensor:

$$\begin{aligned} \Pi_{ij}^{(r)} = & -\left(\frac{1}{5} + \frac{1}{2}C_1 + C_2\right)\left(<u_i u_m>U_{m,j} + <u_j u_m>U_{m,i}\right) + \\ & +\left(\frac{4}{5} - C_1 - \frac{1}{2}C_2\right)\left(<u_i u_m>U_{j,m} + <u_j u_m>U_{i,m}\right) + \\ & +k\cdot(C_1+C_2)(U_{i,j}+U_{j,i}) - (4C_1+C_2)<u_m u_n>U_{m,n}\delta_{ij}. \end{aligned} \qquad (2.10)$$

Since all derivations were made under the assumption of flow incompressibility, the terms in expression (2.9) that contain $\delta_{mn}$ make no contribution in (2.10).

Here, we will emphasize again that although the weak inhomogeneity approximation can be applied to modeling both correlations $<u_i p_{,j}>^{(r)}$ and $<u_{i,j} p>^{(r)}$ (see, e.g., Launder, Reece & Rodi 1975, Speziale, Sarkar & Gatski 1991, and Ristorcelli, Lumley & Abid 1995), Green's theorem cannot be directly used in the analyses of expression (2.6) for the pressure-strain correlation, unless symmetry under permutation of the indices $n$ and $i$ is also assumed. This assumption is, in fact, the assumption of turbulence homogeneity (see Chou 1945, Rotta 1951) and therefore, restricts the application of models using this constrain by homogeneous turbulent flows. Contrary, application of Green's theorem to the analysis of expression (2.5) does not require the assumption of symmetry under permutation of the indices $n$ and $i$. As a result, the application area of model expression (2.10) is extended to inhomogeneous turbulent flows. There is an interesting and useful consequence of that, which will be brought out in relation to expressions (2.10) and (3.1)

Notice also that symmetry under permutation of the indices $n$ and $i$ was imposed on a nonlinear model for the velocity – pressure-gradient correlation suggested in Ristorcelli, Lumley & Abid (1995). As a result, the model suggested there was derived under the assumption of homogeneity and, therefore, is a nonlinear model for the pressure-strain correlation.

## 3. Model coefficients

Let us analyze the constraints on the model coefficients $C_1$ and $C_2$ in expressions (2.9) and (2.10) imposed by limiting states of turbulence.

*Isotropic turbulence.* Setting $<u_i u_j> = 2/3 k \delta_{ij}$, it is easy to show that expression (2.9) satisfies the exact solution for isotropic turbulence subjected to sudden distortion

$$a_{nmji} = k \left( \frac{8}{15} \delta_{ni} \delta_{mj} - \frac{2}{15} \left( \delta_{nm} \delta_{ji} + \delta_{nj} \delta_{mi} \right) \right)$$

(Rotta 1951, Crow 1968, and Reynolds 1976) for any values of the coefficients $C_1$ and $C_2$.

*Homogeneous turbulence.* In the case of homogeneous turbulence, symmetry under permutation of the indices $n$ and $i$ holds for expression (2.8). This condition should be imposed on (2.9). It results in the following relation between two coefficients:

$$\frac{1}{5} - \frac{5}{2}C_1 - C_2 = 0. \tag{3.1}$$

Under condition (3.1), expression (2.9) transforms into

$$\begin{aligned} a_{nmji} &= \left(\frac{4}{5} + C_1\right)<u_i u_n>\delta_{jm} + \left(-\frac{3}{10} + \frac{11}{4}C_1\right)<u_j u_m>\delta_{in} + \\ &+ \left(-\frac{1}{10} - \frac{3}{4}C_1\right)\left(<u_i u_j>\delta_{mn} + <u_i u_m>\delta_{jn} + <u_j u_n>\delta_{im} + <u_m u_n>\delta_{ij}\right) + \\ &+ \left(\frac{1}{5} - \frac{5}{2}C_1\right)k\delta_{in}\delta_{jm} + C_1 k\left(\delta_{ij}\delta_{mn} + \delta_{im}\delta_{jn}\right). \end{aligned} \tag{3.2}$$

Substitution of the coefficient $C_1$ in (3.2) as $C_1 = \frac{6}{55} + \frac{4}{11}C'$ yields the standard linear model suggested by Launder, Reece & Rodi (1975) for the "rapid" part of the pressure-strain correlation with the model coefficient $C'$ (LRR model). This result is expected since in homogeneous turbulence, a model for the velocity – pressure-gradient correlation $<u_i p_{,j}>$ should reduce to a model for the pressure-strain correlation $<u_{i,j} p>$.

We emphasize that the connection between the coefficients $C_1$ and $C_2$ given by expression (3.1) holds only in a homogeneous turbulence, not in general. This yields the LRR model with only a single degree of freedom. But, the coefficients, e.g., $C_1 = 2/5$ and $C_2 = -5/6$ that provide good fits in the wake (see Fig. 1 and discussion below) clearly do not satisfy Eq. (3.1). Thus, there is a benefit to assuming weak inhomogeneity.

*Two-component turbulence*. Let $<\tilde{u}_\alpha^2>$ ($\alpha = 1,2,3$) be the eigenvalues of the Reynolds-stress tensor (i.e., the normal stresses in principal axis) and, for instance, $<\tilde{u}_1^2> = 0$. Then, the following realizability constraint holds: $\tilde{\Pi}_{11}^{(r)} = 0$ (Schumann 1977, Pope 1985, and Shih, Shabbir & Lumley 1994). The sign "~" above a flow variable indicates that this is its value in principal axis of the Reynolds-stress tensor. Using expression (2.10) for $\tilde{\Pi}_{11}^{(r)}$, one obtains

$$\tilde{\Pi}_{11}^{(r)} = 2k(C_1 + C_2)\tilde{U}_{1,1} + (-4C_1 - C_2)\left(<\tilde{u}_2^2>\tilde{U}_{2,2} + <\tilde{u}_3^2>\tilde{U}_{3,3}\right) = 0$$

or

$$C_1 = -C_2 \frac{2k\tilde{U}_{1,1} - \left(<\tilde{u}_2^2>\tilde{U}_{2,2} + <\tilde{u}_3^2>\tilde{U}_{3,3}\right)}{2k\tilde{U}_{1,1} - 4\left(<\tilde{u}_2^2>\tilde{U}_{2,2} + <\tilde{u}_3^2>\tilde{U}_{3,3}\right)}. \quad (3.3)$$

Taking into account that $<\tilde{u}_2^2> + <\tilde{u}_3^2> = 2k$, and $\tilde{U}_{m,m} = U_{m,m} = 0$, expression (3.3) can be rewritten as

$$C_1 = -C_2 \frac{\tilde{U}_{1,1}\left(4k - <\tilde{u}_\beta^2>\right) + 2\tilde{U}_{\beta,\beta}\left(k - <\tilde{u}_\beta^2>\right)}{\tilde{U}_{1,1}\left(10k - 4<\tilde{u}_\beta^2>\right) + 8\tilde{U}_{\beta,\beta}\left(k - <\tilde{u}_\beta^2>\right)}, \quad (3.4)$$

where $\beta = 2$ or 3 (no summation on $\beta$). Again, this connection between coefficients is valid only in the two-component limit.

*Two-component axisymmetric turbulence.* Expression (3.4) reduces to the simple relation between coefficients

$$C_1 = -1/2 \cdot C_2 \quad (3.5)$$

in the case of two-component axisymmetric turbulence, where $<\tilde{u}_2^2> = <\tilde{u}_3^2> = k$.

*Two-component homogeneous turbulence.* Combining expressions (3.1) and (3.4), one can determine the coefficients $C_1$ and $C_2$ in the limit of two-component homogeneous turbulence. The relation between two coefficients is given by (3.1). Substitution of the coefficient $C_2$ found from this expression into (3.4) yields the following expression

$$C_1 = \frac{1}{5} \bigg/ \left(\frac{5}{2} - \frac{\tilde{U}_{1,1}\left(10k - 4<\tilde{u}_\beta^2>\right) + 8\tilde{U}_{\beta,\beta}\left(k - <\tilde{u}_\beta^2>\right)}{\tilde{U}_{1,1}\left(4k - <\tilde{u}_\beta^2>\right) + 2\tilde{U}_{\beta,\beta}\left(k - <\tilde{u}_\beta^2>\right)}\right). \quad (3.6)$$

for the coefficient $C_1$.

*Two-component axisymmetric homogeneous turbulence.* Expression (3.6) reduces to $C_1 = 2/5$ in the case of two-component axisymmetric homogeneous turbulence. The corresponding value of $C_2$ in such a flow is $-4/5$. These values satisfy (3.5).

These examples demonstrate that the coefficients $C_1$ and $C_2$ vary from flow to flow. Except some simple cases where they can take constant values (e.g., the two-component axisymmetric homogeneous turbulence considered here), $C_1$ and $C_2$ are functions of at least the mean velocity gradients and the Reynolds stresses, even in homogeneous turbulence. There are no *universal* constant values for these coefficients. It is true in particular for expression (3.2): this expression can be used to describe the two-component homogeneous turbulence only if the coefficient $C_1$ is given by (3.6), and thus, both homogeneity (3.1) and two-componentiality (3.4) conditions are satisfied. Even though $C_1$ can take a constant value in any given flow, this value varies depending on the flow characteristics.

The conclusion that the coefficients $C_1$ and $C_2$ are functions rather than constants is in agreement with the previous discussion in, e.g., Lumley (1978), Ristorcelli, Lumley & Abid (1995), Reynolds (1987), and Girimaji (2000) who assume that the model coefficients in the pressure-strain correlation models are functions of the mean velocity gradients, Reynolds stresses, and dissipation.

**4. Pressure-diffusion model: "rapid" part**

In the transport equation for the turbulent kinetic energy, expression (2.10) for $\Pi_{ij}^{(r)}$ contracts to a model for the "rapid" part of the pressure diffusion term with only one model coefficient:

$$-\frac{1}{\rho}<u_i p>_{,i}^{(r)} = \left(-\frac{3}{5}+C_k\right)P, \qquad (4.1)$$

where

$$C_k = \frac{15}{2}\cdot C_1 + 3C_2 \qquad (4.2)$$

and $P = 1/2\cdot P_{ii} = -<u_i u_j> U_{i,j}$. In general, the coefficient $C_k$ is a function of the same parameters as the coefficients $C_1$ and $C_2$. In homogeneous turbulence, however, substitution of (3.1) in expression (4.2) yields the universal constant value of $C_k$ equal to $3/5$. That is, in homogeneous turbulence, the pressure diffusion term does not contribute to the turbulent kinetic energy balance as expected. The fact that $C_k$ reduces to the universal constant value in homogeneous turbulence clearly indicates that $C_1$ and $C_2$ should also depend on certain parameters that characterize inhomogeneous effects. A general form of $C_k$ in inhomogeneous turbulence is, however, out of the scope of the current paper.

An important question in modeling $<u_i p>_{,i}$ is whether a model expression for this term should be of the diffusive type. In regard to the "rapid" part of the correlation $<u_i p_{,j}>$ (see expression (2.1)), there is no indication that the model for this term should be of the diffusive type. What "diffusive type" requires is that the integral of the sum of three correlations $<u_i p>_{,i}$ ($i=1,2,3$) taken over the entire flow volume vanishes. It does not imply that the sum of three correlations $<u_i p>_{,i}$ vanishes at every point in the flow, or that any one of $<u_1 p>_{,1}$, $<u_2 p>_{,2}$, and $<u_3 p>_{,3}$ vanishes throughout the flow. This requirement does not also imply that each of the terms in expression (1.3) would vanish separately. In the current paper, only one of the terms in (1.3) is modeled. Finally, even assuming that the integral of (4.1) taken over the entire flow volume should vanish, one can argue that this result can be achieved with different functional forms of the coefficient $C_k$, not necessarily of the diffusive type. This question clearly requires more study in the future. Notice, however, that expressions (2.9), (2.10), and (4.1) hold regardless of the models for the coefficients $C_1$, $C_2$, and $C_k$, and these expressions are the focus of the current paper.

**5. Verification against DNS data**

Even though general mathematical expressions for the coefficients $C_1, C_2$, and $C_k$ are currently unavailable, information on their functional form can be partly drawn from DNS and experimental data. In the present paper, the DNS results for the unforced simulations of the time-developing plane turbulent wake presented in Moser, Rogers & Ewing (1998) are used. The flow was allowed to evolve long enough to attain self-similarity. Therefore, the cross-stream direction $x_2$ is the only direction of flow inhomogeneity and of all mean velocity derivatives, only $U_{1,2} \neq 0$ ($x_1$ is the streamwise direction). The self-similar cross-stream coordinate is defined to be $\xi = x_2/\delta(t)$, where the half-width $\delta$ is the distance between the $x_2$-location at which the mean velocity is half of the maximum magnitude of the velocity deficit.

For $\Pi_{11}^{(r)}$, $\Pi_{33}^{(r)}$, and $\Pi_{12}^{(r)}$, the terms on the right-hand side of expression (2.10) can be evaluated using the DNS data for the Reynolds stresses and $U_{1,2}$. Figure 1 displays the profiles of $\Pi_{11}^{(r)}$, $\Pi_{33}^{(r)}$, and $\Pi_{12}^{(r)}$ (DNS profiles are denoted by dashed lines, and the profiles calculated from expression (2.10) are denoted by solid lines). Interestingly enough, it was found that by assigning to the coefficients $C_1$ and $C_2$ the values 2/5 and -5/6, respectively, one can obtain the profiles for $\Pi_{11}^{(r)}$, $\Pi_{33}^{(r)}$, and $\Pi_{12}^{(r)}$, which are in good agreement with the DNS data. These values of $C_1$ and $C_2$ (which obviously do not satisfy the homogeneity constraint (3.1)) may be considered as a validation of the weak inhomogeneity assumption.

The value of the coefficient $C_k$ ($=0.5$) is obtained from (4.2). In the wake, DNS data are not available separately for the "rapid" and "slow" parts of the pressure diffusion in the transport equation for the turbulent kinetic energy. Therefore, only the sum of expressions (1.2) and (4.1) can be compared with the DNS data. To compute the sum of expressions (1.2) and (4.1), the DNS data for the production and turbulent diffusion terms in the transport equation for the turbulent kinetic energy are used. The result (solid line) is compared with the DNS profile for the pressure diffusion (dotted line) in Fig. 2(a). In addition, the production (dashed line) and the turbulent diffusion (dash-dotted line) are also shown in the figure.

In the self-similar plane turbulent mixing layer, the coefficient $C_k$ can also be approximated by a constant value. The DNS data for this flow are presented in Rogers & Moser (1994). Matching the maximum of the DNS profile for the pressure diffusion with the maximum of the sum of expressions (1.2) and (4.1) calculated with the DNS data for the production and turbulent diffusion terms yields $C_k = 0.52$. The DNS profiles for the pressure diffusion is compared with the calculated sum of expressions (1.2) and (4.1) in Fig. 2(b). Notations are the same as in Fig. 2(a).

As Figures 2(a) and 2(b) demonstrate, the sum of expressions (1.2) and (4.1) closely approximates the DNS profile for the pressure diffusion in both flows. The value of the coefficient $C_k$, which is found to be 0.5 in the wake and 0.52 in the mixing layer, deviates from 0.6, which is the value of $C_k$ in homogeneous turbulence (see discussion in relation to expression (4.2)).

## 6. Summary

In the current paper, a new model expression for the "rapid" part of the velocity – pressure-gradient correlations in inhomogeneous turbulence is presented. Expression (2.10) contains two model coefficients. Generally, they are unknown functions of several parameters including, among other quantities, the mean velocity gradients and the Reynolds stresses. It appears that these coefficients can take constant values or be well approximated by a constant value in some flows. DNS data of Rogers & Moser (1994) and Moser, Rogers & Ewing (1998) are used to determine a value of these coefficients in the self-similar plane wake and the self-similar plane mixing layer. However, even in flows with simple geometries, such as the wake and the mixing layer considered here, the turbulence structure is complicated. As discussed in Rogers & Moser (1994) and Moser, Rogers & Ewing (1998), both experiments and DNS show that in geometrically-equivalent flow situations at the same Reynolds number, multiple asymptotic states can be observed. The difference between alternative states manifests itself in statistics and the flow structure. Whereas the mean velocity and shear stress profiles are universal (or nearly universal) under appropriate scaling, the normal stresses and turbulent kinetic energy profiles are non-unique. A model with constant-value coefficients cannot describe this phenomenon. DNS confirms that statistical differences reflect the differences in the large-scale structure of turbulence, which depends strongly on the Reynolds number, "uncontrolled and possibly unknown properties of the initial or inlet conditions" (Moser, Rogers & Ewing 1998), flow geometry, boundary conditions, external forces etc. (Tsinober 1998). Further study (including DNS) is necessary to determine how the large-scale structure of turbulence is reflected in the model coefficients and how they can be described mathematically in the general case.

**Acknowledgments**

A part of the work was conducted when the author was affiliated with the Center for Turbulence Research (Stanford University) and ONERA-Toulouse, France. The author would like to thank Michael Rogers (NASA-Ames) for providing the DNS data and M. Y. Hussaini for permission and support for the preparation of this work for publication.

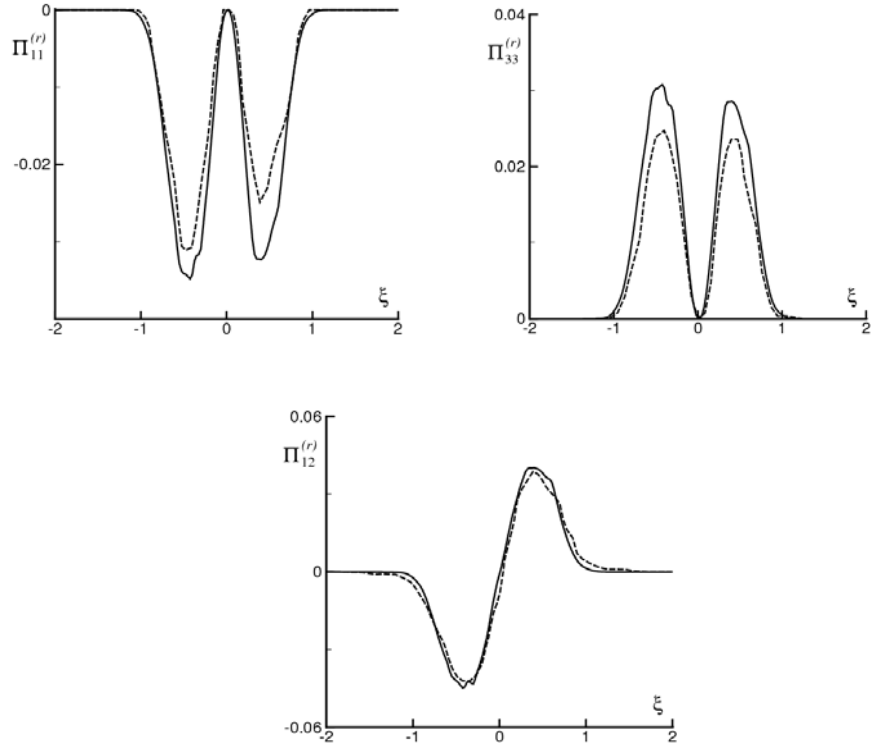

**Figure 1** "Rapid" part of the velocity – pressure-gradient correlations in the Reynolds-stress budget in the wake. Notations: DNS profiles (dashed lines), profiles calculated from expression (2.10) (solid lines).

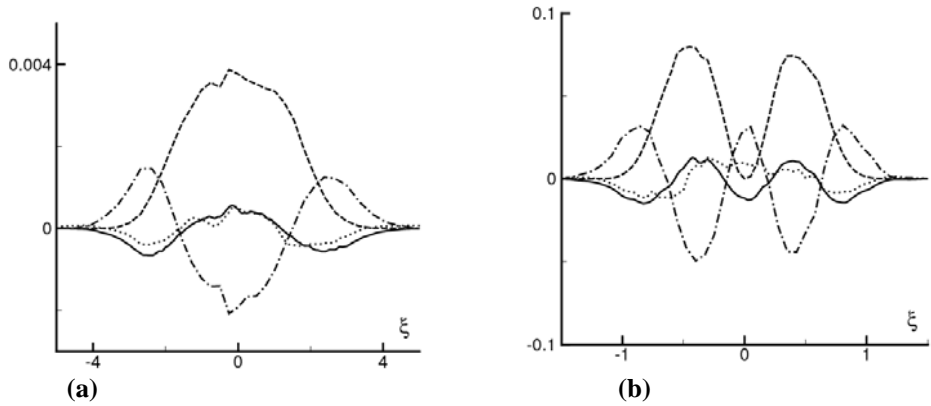

**Figure 2** Partial turbulent kinetic energy balance: (a) wake, (b) mixing layer. Notations: calculated sum of expressions (1.2) and (4.1) (solid lines); DNS pressure diffusion (dotted lines); DNS production (dashed lines), and DNS turbulent diffusion (dash-dotted lines).

15